\documentclass[12pt,preprint]{aastex}
\slugcomment{draft version}

\begin{document}

\title{A Cr-K emission line survey in young supernova remnants with {\sl Chandra}}
\author{X.J. Yang\altaffilmark{1,2}, H.Tsunemi\altaffilmark{3}, F.J. Lu\altaffilmark{1},
 L. Chen\altaffilmark{2}}

\altaffiltext{1}{Particle Astrophysics Center, Institute of High Energy Physics,
Chinese Academy of Sciences, Beijing 100049, P.R. China, yangxj@mail.ihep.ac.cn,
lufj@mail.ihep.ac.cn}
\altaffiltext{2}{Department of Astronomy, Beijing Normal University, Beijing
100875, P.R. China}
\altaffiltext{3}{Department of Earth and Space Science, Graduate School of Science, Osaka
University, Toyonaka, Osaka 560-0043, Japan; tsunemi@ess.sci.osaka-u.ac.jp}

\begin{abstract}

We performed a Cr-K emission line survey in young supernova remnants (SNRs)
 with the {\sl Chandra} archival data. Our sample includes
W49B, Cas A, Tycho and Kepler. We confirmed the existence of the Cr line
in W49B and discovered this emission line in the other three SNRs.
The line center energies, equivalent widths (EWs) and fluxes of the Cr lines
are given. The Cr in Cas A is in a high ionization state while that in
Tycho and Kepler is in a much lower one. We find a good positive correlation
between Cr and Fe line center energies, suggesting a common origin
of Cr and Fe in the nucleosynthesis, which is consistent with the
theoretical predictions. We propose that the EW ratio between Cr and
Fe can be used as a supplementary constraint on the progenitors'
properties and the explosion mechanism.

\end{abstract}


\section{Introduction}
The X-ray emission of young supernova remnants (SNRs) is predominantly from
the ejecta heated by the reverse shock. Since the ejecta are metal abundant,
their X-ray spectra usually show strong emission lines of heavy elements such as
O, Ne, Mg, Si, S, Ar, Ca and Fe. Many works have focused on
these emission lines from young SNRs, e.g. W49B (Hwang et al. 2000a; Miceli et al.
2006), Tycho (Tsunemi et al. 1986; Hwang et al. 1997), Cas A
(Tsunemi et al. 1986; Holt et al. 1994; Hughes et al. 2000; Hwang
et al. 2000b) and Kepler (Hatsukade et al. 1990; Kinugasa \&
Tsunemi, 1999; Cassam-Chena\"i et al. 2004; Reynolds et al. 2007).
Comparision of the abundance pattern of an SNR with the theoretical
predictions could be used to constrain the properties of the corresponding
supernova (SN) (e.g., Hughes et al. 1995; Rakowski et al. 2006).

For a long time the K lines from heavy elements such as Ti, Cr and
Mn, which are expected to appear in 4$-$6 keV had not been detected
in SNRs. The first and the only case, as far as we know, is the detection of
the He-like lines of Cr and Mn in SNR W49B by the Advanced Satellite for Cosmology and
Astrophysics ({\sl ASCA}) (Hwang et al. 2000a) and {\sl XMM-Newton} (Miceli et al. 2006).
Hwang et al. (2000a) proposed that Cr, Mn and Ni would be the most promising
heavy atomic species for future detection, since they are the most
abundant elements next to Fe with K-shell emission lines at energies
above $\sim$4 keV.

From nucleosynthesis theories (Woosley \& Weaver 1994; Woosley
\& Weaver 1995; Thielemann et al. 1996) we know that Cr is mainly
formed from the decay of Fe during SN explosions. SNRs from type Ia
explosions are the top candidates for the detection of Cr emission, as the
progenitors are completely destroyed during the SN explosions and
all material is ejected into interstellar space. The Cr line emission could
also be detected in a core-collapse SNR, if there is a large
fraction of Cr formed outside the mass cut. In this paper, we report
a Cr line survey in the young and bright SNRs W49B, Cas A, Tycho and
Kepler using the abundant archival data collected by the {\sl Chandra}
X-ray observatory. We describe the observations and data analyses in
$\S$ 2. The results are shown in $\S$ 3, discussion in $\S$ 4
and summary in $\S$ 5. All through this paper, the statistical
uncertainties are given at 90\% confidence level.

\section{Observations and Data analyses}

SNRs are important observational targets for {\sl Chandra}. For
example, as one of the largest projects, {\sl Chandra} has observed the Cas A
SNR for a few times with a total exposure time
of 1 Ms (Hwang et al. 2004). The Kepler SNR was also observed for 750 ks (Reynolds et al. 2007).
In this paper, we collected almost all the available {\sl Chandra} ACIS-S
data of the four SNRs. The detailed observation information is listed
in Table 1. The data were processed using the CIAO software package
(version 3.4). We created new level 2 event files for all the
observations we used, applying gain map, time-dependent gain and
charge transfer inefficiency corrections with the latest released
calibration files (CALDB version 3.4.3 and ATOMDB version
1.3.1). The only exception is that the charge transfer inefficiency can
not be corrected for all the Cas A data since
they are acquired in GRADED mode. Fig. 1 shows the X-ray
images of the four SNRs. Since the Cr line is very weak,
we extracted the spectrum of each SNR from almost the entire source region, as
illustrated in Fig. 1. The background spectra were extracted from
the off-source regions.

Cas A and Kepler have been observed for a few times spanning 4 to 5 months.
We therefore carefully performed the analysis to eliminate the
effect of any CCD performance evolution with time.
Taking Cas A as an example, we first created source and background
spectra as well as the corresponding RMF and ARF files for each
of the 9 observations (including IDs 4634, 4635, 4636, 4637, 4638, 4639, 5196,
5319, 5320), and then combined them with FTOOLs
{\footnote{http://heasarc.gsfc.nasa.gov/docs/software/ftools/}}.
This process has taken into account the degradation of the CCD
performance during the observation span.  We noticed that
there is no clear difference between the 9 RMF files, therefore the
ARF files can be added with FTOOLs {\sl addarf}. The total photon
number of each source spectrum was taken as the corresponding adding
weight. We performed a joint fit to the spectra of all the observations
by using their respective RMF and ARF files, and found
that the fitting results are consistent with those obtained by
using the combined data and the added RMF and ARF. This demonstrates
that the combining process works well and
suggests that the gain is properly calibrated from observation to
observation. The spectral fitting was done with XSPEC version 11.3.2
(Arnaud 1996).

The 0.5$-$8.0 keV spectra of the four SNRs are plotted in Fig. 2.
Apparently the spectrum of Cas A has the best statistics, therefore
we take it as an example to describe our spectral analysis process. We first
fitted the continuum-dominant 4.2$-$6.0 keV spectrum with bremsstrahlung
and power law models respectively. In both cases, three line-like features
appear in the residual distribution. In this case, we added three Gaussian
components to account for these lines.
The fitting models and residuals are plotted
in Fig. 3. The best fit centroid energies of the three lines are
4.607, 4.878 and 5.635 keV respectively. The power law model can
represent the continuum emission better than the bremsstrahlung one,
with $\chi^2$ 452.9 vs. 560.3 for 111 degrees of freedom (d.o.f.),
although neither of the two fits is statistically acceptable.
However, adapting different continuum models does not change
the best fit results of the line strength significantly, e.g. the equivalent
width (EW) of the 5.635 keV line is 9.6$\pm{0.87}$ eV with the power law model
and 10.7$\pm{0.95}$ eV with the bremsstrahlung. For the other three
SNRs, the power law model also fits the continuum emission better
than the bremsstrahlung model does. The latter model seems
to overestimate the spectra at low energies, while underestimating
at high energies. In this case, the power law model was used in
the following fitting process to account for the continuum emission.
We also note here that the 4.2$-$6.0 keV spectrum of Cas A is more
complicated than the model we used. This will be further discussed in $\S$ 3.2.

According to the Astrophysical Plasma Emission Code
(APEC{\footnote{http://cxc.harvard.edu/atomdb}}), the two lines
below 5.0 keV could be the He-like Ca-K emission, while the Ti-K emission
could not be ruled out as it also emits around those energies.
Unfortunately, Ti has not been considered in the currently available plasma codes,
thus we cannot get the emissivities of the Ti lines to judge whether they
represent these observed line features or not. Since the
main topic of the current paper is Cr emission and the
line candidates below 5.0 keV are beyond its scope, we only adopt
data above 5.0 keV for the analyses of the continuum and line
emission. As an element of the Fe-group, Cr is closely related to Fe in its
synthesis process (Woosley \& Weaver 1994; Woosley \& Weaver 1995;
Thielemann et al. 1996). A comparative study of the Fe and Cr
emission should be important. Therefore we take the 5.0$-$7.5 keV
spectra to study both the Cr and the Fe K lines in the following.

We note here that the systematic errors of the EWs and fluxes induced by
using different continuum models and different energy ranges are included in
the parameters' confidence ranges in Tables 2 \& 3.
By fitting the 5.0$-$7.5 keV spectra of the four SNRs with bremsstrahlung
to represent the continuum model, we can get the best fit EWs and fluxes
of the Cr and Fe emission lines along with their 90\% statistical errors.
Likewise, by fitting these spectra with the continuum energy extended down to
4.2 keV, we can also get another set of best fit EWs, fluxes and their confidence ranges.
The final confidence ranges listed in Tables 2 \& 3 cover those
from the above two fittings and from fitting with a power law continuum model
to the data in 5.0$-$7.5 keV.

\section{Results}

\subsection {W49B}

In the analysis of the {\sl ASCA} spectrum, Hwang et al. (2000a)
found evidence of the He-like Cr and Mn lines near 5.69 and 6.18
keV, and the existence of these lines was confirmed by {\sl XMM-Newton}
(Miceli et al. 2006). From the
spectra (Fig. 2 (top left) and Fig. 4 (top left)), we can see that
the Cr and Mn lines are also detected with {\sl Chandra}. We fitted
the 5.0$-$7.5 keV spectrum with a power law model plus three
Gaussian components to account for the continuum, Cr, Mn and Fe
lines respectively. Eliminating the Gaussian component for Cr (or Mn)
leads to an increasing of $\chi^2$, from 93.4 for d.o.f
68 to 185.8 (or 103.7) for d.o.f
71. The Cr line emission is firmly detected in the {\sl Chandra}
spectrum, while the Mn line is less significant.

Fig. 4 (top left) shows the fitting model and residuals, and Table 2
gives the best fit results for the Cr and Fe lines, including line
center energies and fluxes along with their uncertainties. The EWs
of the two lines, calculated from the XSPEC command {\sl eqwidth},
are given in the same table. From our fitting analyses, the Cr line
center energy is $E_{Cr}$ = 5.656$^{+0.014}_{-0.016}$ keV, and the
flux is $f_{Cr}$ = (0.32$^{+0.08}_{-0.07}$) $\times 10^{-4}$ photons
cm$^{-2}$ s$^{-1}$, while for the Mn line, $E_{Mn}$ =
6.126$^{+0.035}_{-0.030}$ keV and $f_{Mn}$ = (0.10$\pm{0.05}$)
$\times 10^{-4}$ photons cm$^{-2}$ s$^{-1}$. The fitted line
parameters are generally in good agreement with those from {\sl
ASCA} and {\sl XMM-Newton} (c.f Table 3).

\subsection {Cas A}

The 5.0$-$7.5 keV spectrum of Cas A is shown in Fig. 4 (top right).
It was fitted with a power law (for continuum) plus three Gaussian
components (for the lines). Two Gaussian components were used to
represent the Fe-K complex since it cannot be well fitted by
a single Gaussian profile. The centroid energies for these two Gaussian
components are 6.600$\pm{0.001}$ keV and 6.659$\pm{0.001}$ keV, with their
widths 100$\pm{2}$ eV and 72$\pm{7}$ eV respectively.  The centroid energy
of the Fe-K complex listed in Table 2 is the emission measure weighted
mean value, while the flux and EW are the sums of the two components.
The third Gaussian component is for the 5.635 keV line mentioned in $\S$ 2.
Adding the third Gaussian leads to a much better
fit (c.f. Fig. 3), with the reduced $\chi^{2}$ decreasing from
993.8/65 to 222.9/62 for data in 5.0$-$6.0 keV. This represents a
significant detection of the Cr-K line emission, and its best fit parameters
are listed in Table 2 as well.

The requirement of two Gaussian components for the Fe-K complex is
probably due to the Doppler shift variation across the whole remnant. Spatially
resolved X-ray spectroscopy of Cas A by both {\sl XMM-Newton}
(Willingale et al. 2002) and {\sl Chandra} (Yang et al. 2008) shows
that different parts of this remnant are moving with different
line of sight velocities, with typical value of $\pm$1000 km
s$^{-1}$. Considering this, we selected two regions based on the
Doppler shift map from the {\sl XMM-Newton} observation (Willingale
et al. 2002), as illustrated in Fig.1. The spectra of the two
regions are given in Fig. 5.  They were fitted with a power
law plus two Gaussian components, representing the continuum, Cr and Fe-K line emission.
The Cr line is very significant in both spectra. Adding a Gaussian
component for Cr leads to the reduced $\chi^{2}$ decreasing from 440.0/62 to 197.4/59
and from 670.9/62 to 184.2/59 for the blueshifted
and redshifted regions respectively in 5.0$-$6.0 keV .
The fitted line center energies of the Fe-K lines are
6.591$\pm{0.001}$ keV and 6.669$\pm{0.001}$ keV, which
are generally consistent with the line center energies of the
two components fitted to the Fe-K line of the whole remnant spectrum.
This demonstrates that the
Doppler shift dominates the structure of the Fe-K complex of Cas A.
The centroid energies of the Cr line in the two regions are
5.590$^{+0.008}_{-0.007}$ and 5.657$^{+0.010}_{-0.008}$ keV, with
a separation similar to that of the two Fe-K lines. The Cr and Fe
ejecta are thus probably moving with similar velocity.
We noticed that the Fe-K line from a small region of Cas A also
cannot be well represented by one Gaussian component (c.f Fig. 5,
residual distribution), suggesting small velocity
difference even within such a scale (c.f Fig. 7 in Willingale et al.
2002).

On the other hand, the existence of more than one strong Fe-K line
that cannot be resolved with {\sl Chandra} ACIS could also
contribute to the broadening of the Fe-K line in Cas A.
Various Fe-K lines around this energy have been clearly displayed in
the spectra of the cataclysmic variables V1223 Sagittarii (Mukai et
al. 2001) and U Geminorum (Szkody et al. 2002) using data collected
by the High Energy Transmission Grating onboard {\sl Chandra}. The
centroid energy difference of the two Gaussian components in Cas A is
about 60 eV, which is comparable with the difference of the He-like
triplet of Fe-K lines ($\sim$65 eV {\footnote{c.f Mewe et al. 1985;
http:
//www.camdb.ac.cn/e/spectra/spectra\_{}search.asp}}). Higher energy resolution
spectra from future missions will help us to resolve the Fe-K
complex.

\subsection {Tycho \& Kepler}

The 5.0$-$7.5 keV spectra of Tycho and Kepler are shown in Fig.
4. We first fitted the spectrum of Tycho with a power law (for
continuum emission) plus one Gaussian component (for the Fe-K line). A
line-like structure came out around 5.45 keV in the residual
distribution. As the neutral Cr K line is near 5.4 keV while the
H-like Cr K line is around 5.9 keV (c.f Table 4), this structure
might be the Cr line in relatively low ionization state. In this
case, we added a Gaussian component there, and the reduced $\chi^{2}$
decreased from 36.6/21 to 11.4/18 in 5.0$-$6.0 keV.
For the Kepler spectrum in the same energy range, adding a Gaussian
component around 5.46 keV also leads to a reduction of the reduced
$\chi^{2}$ from 47.6/24 to 27.8/21. Finally, the 5.0$-$7.5 keV spectra of
Tycho and Kepler were fitted with a power law (for continuum)
plus two Gaussian components (one for the Cr line and the other for the
Fe line). The fitting models and residual distribution are also
given in Fig. 4.

The best fit parameters for the Cr and Fe lines are listed in Table
2, along with their confidence ranges. Although there are big
uncertainties in the EWs and fluxes of Cr lines in both Tycho and
Kepler, the detection of these lines is significant, judging from
the changes of $\chi^{2}$ mentioned above and the lower limits of
the confidence ranges. From Table 2, we can see that the centroid
energies of the putative Cr lines in Tycho and Kepler are about 200
eV lower than those of W49B and Cas A. This will be further
discussed in $\S$ 4.1.

\section{Discussion}

\subsection{The ionization state and spatial correlations between Cr and Fe}

The different line center energies of the Cr emission lines in these
SNRs reflects that the ionization states of Cr in these SNRs are dissimilar.
The Cr in W49B is in relatively high ionization state since its line
center energy is 5.656$^{+0.014}_{-0.016}$ keV (c.f Table 2). For
Cas A, the centroid energy of the Cr line is
5.635$^{+0.007}_{-0.005}$ keV, which also suggests a high
ionization state of Cr. The center energies of the putative Cr lines
in Tycho and Kepler are about 200 eV lower than those in W49B and
Cas A. If these lines do come from Cr, it should be in a relatively
low ionization state (c.f Table 4). We noticed that the Fe-K line
center energies in Tycho and Kepler are both around 6.44 keV. This
implies low ionization states of Fe as well, as suggested by previous
observations (Tsunemi et al. 1986;
Hwang et al. 1998; Hwang et al. 2002; Kinugasa \& Tsunemi 1999).

In Fig. 6, we plot the Cr line center energy versus that of the Fe line
in these four SNRs. The theoretical centroid energies of Cr and Fe K
lines in different ionization states given in Table 4 are
overplotted in the same figure. Obviously, there is a positive
correlation between the two line center energies both theoretically
and observationally. We can conclude from Fig. 6 that the emission
lines around 5.46 keV in Tycho and Kepler are from Cr. Meanwhile,
the Cr in Cas A and W49B might be He/Li-like, while Ne-like or an even
lower ionization state in Tycho and Kepler.
Since the line center energy is closely related to the ionization
ages of the emitting plasma, such a positive correlation implies
that the ionization ages of Cr and Fe are closely related to each
other.

The above ionization state correlation suggests that the Cr and
Fe ejecta are co-located, which is also supported by the spectra
of the blueshift and redshift regions in Cas A. As shown in $\S$ 3.2,
the Cr and Fe ejecta are probably moving with the same velocity as implied by
the centroid energies of the Cr and Fe lines from these two regions.
Meanwhile, the EW ratio of the Cr line to the Fe line in the blueshift
region spectrum is 0.82\%, while it is 0.91\% in the redshift one. They are both
consistent with the overall value (0.85$^{+0.18}_{-0.07}$\%, c.f. Table 2) within the
confidence range. This is further evidence that Cr and Fe are co-located.

According to the nucleosynthesis theory (Woosley \& Weaver
1994; Woosley et al. 1995), $^{50}$Cr is generated by explosive
oxygen and silicon burning, while $^{52,53}$Cr are the products of
$^{52,53}$Fe decay in explosive silicon burning. The most abundant
Fe is $^{56}$Fe decayed from $^{56}$Ni, which is also
produced mainly from explosive silicon burning. Among all the
isotopes of Cr, $^{52}$Cr is the most abundant, and its production is
generally at least one order of magnitude greater than the others
(Nomoto et al. 1984; Woosley et al. 1995). In this case, most of the Cr
would be located near and thus share a similar ionization time
with Fe in SNRs.
Cayrel et al. (2004) observed a number of so-called ``first stars'',
i.e. very metal-poor dwarfs and giants. They found that the scatter of the
Cr/Fe values of these stars is very small, indicating that the
production of Fe and Cr are very closely linked. Our results are consistent
with the theoretical predictions and further strengthen the previous
observational statement in a different way.

\subsection{Using the Cr to Fe EW ratio to constrain the SN explosion process}

Many theoretical calculations of nucleosynthesis have included Cr
for both Ia (Woosley \& Weaver 1994; Iwamoto et al. 1999) and
core-collapse SNe (Woosley \& Weaver 1995; Thielemann et al. 1996;
Maeda \& Nomoto 2003). In order to compare with these models, we
need the mass (or abundance) ratio between Cr and the other
elements. Hwang et al. (2000a) interpolated the He$\alpha$
emissivity of Cr from those of Si, S, Ar, Ca and Fe, using the
Raymond-Smith (RS) code for the 2 keV plasma in collisional ionization
equilibrium (CIE), and then calculated the Cr abundance of
W49B. They found that the Cr and Fe abundances are consistent with a
solar ratio, corresponding to an atomic number ratio $1.0\%$ of Cr
to Fe (Anders \& Grevesse 1989) and thus a mass ratio $M_{Cr/Fe}$
$\sim0.9\%$. Cr and Fe in Tycho and Kepler are in relatively low
ionization state, and as emissivities in these states are not well calculated for
elements such as Si, S, Ar, Ca, and Fe, it is difficult to
interpolate the Cr emissivity as done by Hwang et al. (2000a).
However, according to the discussion in $\S$4.1, Cr and Fe are
spatially correlated and thus have similar temperatures and
ionization ages, therefore the EW ratio of the Cr and Fe K lines
(EW$_{Cr/Fe}$) would be a good representation of the mass ratio of
these two elements. Considering this, we use the EW ratio for the
discussion below.

From Table 2, we can see that EW$_{Cr/Fe}$ of these four SNRs differ from
one another. This ratio might be used to constrain the properties of
the corresponding SNe. Badenes et al. (2006) have made detailed
comparisons of the X-ray spectra of the type Ia SNR Tycho with the
theoretical models. They found the one-dimensional delayed
detonation model can well reproduce its X-ray emission. From
numerical calculations, the standard SNe Ia models, i.e carbon
deflagration and Chandrasekar mass models (e.g W7, W70 etc, Nomoto
et al. 1984) often yield relatively small $M_{Cr/Fe}$ ($<1\%$,
Nomoto et al. 1997; Iwamoto et al. 1999). Multi-dimension models
based on W7 also give similar $M_{Cr/Fe}$ (Travaglio et al. 2004;
2005). On the other hand, the delayed detonation models (WDD, CDD
etc) produce much larger $M_{Cr/Fe}$ ($> 2\%$), and $M_{Cr/Fe}$ decreases as
the transition density increases (Nomoto et al. 1997;
Iwamoto et al. 1999). Therefore, our
observational results of Tycho suggest that there should be a
deflagration-detonation transition at some stage of Tycho's SN
explosion, which further confirm the results of Badenes et al.
(2006). Meanwhile, we favor a relatively small transition density,
probably $1.7\times10^7$ g cm$^{-3}$ (Nomoto et al. 1997; Iwamoto et
al. 1999). This is also consistent with that suggested by
Badenes et al. (2006, $2.2\times10^7$ g cm$^{-3}$).

Cas A has been identified as the remnant of a core-collapse SN. The
recent study of its progenitor implies the mass to be 15$-$25
$M_{\odot}$ (Young et al. 2006). According to the calculations of
spherical models by Thielemann et al. (1996), a 20$M_{\odot}$
progenitor gives a $M_{Cr/Fe}$ of 1.2\%, which is higher than our
measured value. However the non-spherical explosion may lead to a
smaller $M_{Cr/Fe}$ (Maeda \& Nomoto 2003). It has already been
suggested that the explosion of Cas A is asymmetric, based on the
jet structure (Hwang et al. 2004) and the Doppler map (Willingale et
al. 2002). Meanwhile, a bigger progenitor mass would lead to a smaller
mass cut and thus smaller $M_{Cr/Fe}$, since Cr is
mainly produced in the incomplete Si-burning zone (Umeda \& Nomoto 2002).
Therefore, we support a higher progenitor mass (Young et al. 2006)
and the asymmetric explosion scenario for Cas A.

The classifications of W49B and Kepler are not conclusive. For W49B,
Hwang et al. (2000a) compared the relative abundances of Mg, S, Ar,
Ca, Fe and Ni to Si, and suggested that W49B may have a type Ia
progenitor. However they claimed that a low mass (13$-$15
$M_{\odot}$) type II progenitor is also possible. The {\sl Chandra}
image of W49B shows a bipolar jet, which was taken as evidence for a
gamma-ray burst (GRB)
remnant{\footnote{Http://chandra.harvard.edu/press/04\_{}releases/press\_{}060204.html}}.
This was further supported by multi-band observations (Keohane
et al. 2007). The nucleosynthesis calculation for bipolar
core-collapse SN explosions (Maeda \& Nomoto 2003) generally
predicts an EW$_{Cr/Fe}$ of $1.0\pm0.5$\%, which matches the
observations perfectly (Hwang et al. 2000a; Miceli et al. 2006; this
paper). Kepler tends to be identified as a type Ia SNR (Reynolds et al. 2007).
If so, the small EW$_{Cr/Fe}$ would favor the carbon deflagration
models (W7, W70 etc) rather than those involving detonation (Nomoto
et al. 1997; Iwamoto et al. 1999; Travaglio et al. 2004;
2005).

The above discussions are based on the overall mass ratio of Cr to
Fe in an SNR. It is possible that in these SNRs there are still a
fraction of Fe and Cr that has not been overtaken by the reverse
shock and so invisible in X-rays. However, according to our
discussion in $\S$4.1, Cr and Fe should be well mixed, so the
observed EW ratio of Cr to Fe could represent the overall Cr to Fe
mass properly. Therefore the main conclusions are reliable no matter
what fraction of Fe is observed.

\section{Summary}

We performed a Cr K line survey with the {\sl Chandra} data in
young SNRs W49B, Cas A, Tycho and W49B. We confirmed the Cr line in
W49B, and gave a consistent flux and line center energy with respect
to the previous results. Then we report, for the first time, the
detection of Cr lines in Cas A, Tycho and Kepler. We conclude that Cr
in Cas A is in a high ionization state similar to that of W49B, while
Cr is in a low ionization state low in Tycho and Kepler. We find that Cr and Fe have similar
ionization states and are co-located in these four SNRs. The reason
might be that Cr and Fe are synthesized by the same process deep inside the progenitor.
We propose that the EW ratio of Cr to Fe might be used as an supplementary
constraint on the properties of the SN explosions. For the type Ia SNR Tycho,
EW$_{Cr/Fe}$ favors the delayed detonation model with relatively small transition
density ($1.7\times10^7$ g cm$^{-3}$) from deflagration to
detonation. This is consistent with the model suggested by Badenes
et al. (2006) from the comparison of Tycho's X-ray spectra with
theoretical calculations. The relatively small EW$_{Cr/Fe}$ in Cas A
and W49B suggests their asymmetric explosions, which is also
consistent with the previous results. If we adopt the type Ia origin for
Kepler, its small EW$_{Cr/Fe}$ could be attributed to the carbon
deflagration explosion.

\acknowledgements
We are grateful to the anonymous referee for very helpful comments and
suggestions leading to significant improvements of the paper.
The manuscript is read by Dr. E. Miller in MIT.
This work is supported by the Nature Science Foundation of
China through grants 10533020, 10573017, 10778716 and
by a Grant-in-Aid for Scientific Research by the Ministry
of Education, Culture, Sports, Science and Technology, Japan(16002004).

\clearpage

\begin{table*}
\caption[]{Information of {\sl Chandra} observations we used}
\label{obsinfor_02}
\begin{tabular}{cccc}
\noalign{\smallskip} \hline \hline \noalign{\smallskip}
  Target      &   Obs$\_$ID     &  $t_{exp} (ks)$  &    Obs-date             \\
\noalign{\smallskip} \hline \noalign{\smallskip}
  W49B        &     117         &  $\sim$50        &    July, 2000                \\ \hline
  Cas A       &   VLP\tablenotemark{\ast}  &  $\sim$1000      &    February$\sim$May, 2004   \\ \hline
  Tycho       &     115         &  $\sim$50        &    September, 2000           \\ \hline
  Kepler      &  LP\tablenotemark{\dagger} &  $\sim$750       &    April$\sim$August, 2006   \\ \hline

\noalign{\smallskip} \hline \noalign{\smallskip}

\end{tabular}

\tablenotetext{\ast}{Very Large Project, Observation IDs: 4634, 4635, 4636, 4637, 4638, 4639, 5196, 5319 and 5320.}
\tablenotetext{\dagger}{Large Project, Observation IDs: 6714, 6715, 6716, 6717 and 6718.}

\end{table*}

\begin{table*}
\caption{The Cr and Fe line parameters of W49B, Cas A, Tycho and Kepler}
\label{tychocas_01}
\begin{tabular}{cccccccc}
\noalign{\smallskip} \hline \hline \noalign{\smallskip}
  & \multicolumn{3}{c}{Cr} & \multicolumn{3}{c}{Fe}    &      \\
\cline{2-4}  \cline{5-7}   \\
SNR & line center & EW & flux & line center & EW & flux  & EW$_{Cr/Fe}$  \\
 & eV & eV & $\frac{10^{-4}photons}{cm^{2} s}$ & eV & eV & $\frac{10^{-4}photons}{cm^{2}s}$  &   \\
\noalign{\smallskip} \hline \noalign{\smallskip}
W49B   &  5656$^{+14}_{-16}$   &    79.6$^{+20.3}_{-18.2}$  &     0.32$^{+0.08}_{-0.07}$     &   6666$^{+1}_{-2}$   &   5000$^{+280}_{-250}$  &   10.2$^{+0.6}_{-0.5}$     & 1.6$^{+0.40}_{-0.36}$\%    \\
CasA   &  5635$^{+7}_{-5}$     &    8.31$^{+1.72}_{-0.69}$  &     0.65$^{+0.13}_{-0.05}$     &   6633$^{+1}_{-1}$   &   981$^{+12}_{-10}$     &   48.3$^{+0.03}_{-0.03}$   & 0.85$^{+0.18}_{-0.07}$\%   \\
Tycho  &  5465$^{+45}_{-50}$   &    46.2$^{+40.8}_{-14.6}$  &     0.33$^{+0.29}_{-0.10}$     &   6436$^{+6}_{-7}$   &   1280$^{+52}_{-42}$    &   6.13$^{+0.24}_{-0.22}$   & 3.6$^{+3.2}_{-1.1}$\%    \\
Kepler &  5469$^{+40}_{-45}$   &    16.5$^{+8.0}_{-6.6}$    &     0.03$^{+0.014}_{-0.012}$   &   6443$^{+1}_{-2}$   &   2630$^{+80}_{-30}$    &   3.61$^{+0.05}_{-0.04}$   & 0.63$^{+0.30}_{-0.25}$\%   \\

\hline \noalign{\smallskip} \hline \noalign{\smallskip}

\end{tabular}
\end{table*}

\begin{table*}
\caption{The Cr, Mn and Fe line parameters of W49B from different instruments.}
\label{tychocas_02}
\begin{tabular}{ccccccc}
\noalign{\smallskip} \hline \hline \noalign{\smallskip}
  & \multicolumn{2}{c}{Cr} & \multicolumn{2}{c}{Mn}  & \multicolumn{2}{c}{Fe}             \\
  \cline{2-3}  \cline{4-5}  \cline{6-7} \\
Instrument  &   line center   &  flux   &  line center  &  flux  &  line center  &  flux  \\
 & eV & $\frac{10^{-4}photons}{cm^{2} s}$ & eV & $\frac{10^{-4}photons}{cm^{2}s}$ & eV
 & $\frac{10^{-4}photons}{cm^{2}s}$  \\
\noalign{\smallskip} \hline \noalign{\smallskip}
{\sl ASCA}\tablenotemark{a}       &  5685$^{+20}_{-27}$  & 0.30$^{+0.08}_{-0.11}$  &  6172$^{+47}_{-49}$ & 0.13$^{+0.14}_{-0.06}$  &   6658$^{+3}_{-2}$   & 8.30$\pm 0.30$       \\
{\sl XMM-Newton}\tablenotemark{b} &  5660$\pm{10}$       & 0.25$\pm{0.04}$         &  6170$\pm{50}$      & 0.10$\pm{0.03}$         &    $-$            &     $-$                 \\
{\sl Chandra}\tablenotemark{c}    &  5656$^{+14}_{-16}$  & 0.32$^{+0.08}_{-0.07}$  &  6126$^{+35}_{-30}$ & 0.10$\pm{0.05}$         &   6666$^{+1}_{-2}$   & 10.2$^{+0.6}_{-0.5}$ \\
\hline \noalign{\smallskip} \hline \noalign{\smallskip}
\end{tabular}

\tablenotetext{a}{Hwang et al. 2000a}
\tablenotetext{b}{Miceli et al. 2006}
\tablenotetext{c}{this paper}
\end{table*}

\begin{table*}
\caption[]{The K shell line energies of Cr and Fe in different
ionization states.} \label{obsinfor_01}
\begin{tabular}{ccccccc}
\noalign{\smallskip} \hline \hline \noalign{\smallskip}
  element      &   neutral\tablenotemark{a}    &   B-like\tablenotemark{b}    &   Be-like\tablenotemark{b}    &     Li-like\tablenotemark{b}    &     He-like\tablenotemark{b}    &      H-like\tablenotemark{c}      \\
\noalign{\smallskip} \hline \noalign{\smallskip}
Cr\tablenotemark{\dagger}   &  5.420   &   5.533   &   5.580   &   5.657    &     5.665  &   5.916     \\
Fe\tablenotemark{\ddagger}  &  6.405   &   6.549   &   6.617   &   6.652    &     6.674  &   6.966     \\
 \hline \noalign{\smallskip} \hline \noalign{\smallskip}
\end{tabular}

\tablenotetext{a}{K$\alpha$ line energy}
\tablenotetext{b}{line energy with transition from level 2p to 1s}
\tablenotetext{c}{Ly$\alpha$ line energy}
\tablenotetext{\dagger}{http://www.camdb.ac.cn/e/spectra/spectra\_{}search.asp}
\tablenotetext{\ddagger}{Mewe et al. 1985}
\end{table*}

\clearpage

\begin{figure*}
\includegraphics[width=\textwidth,clip]{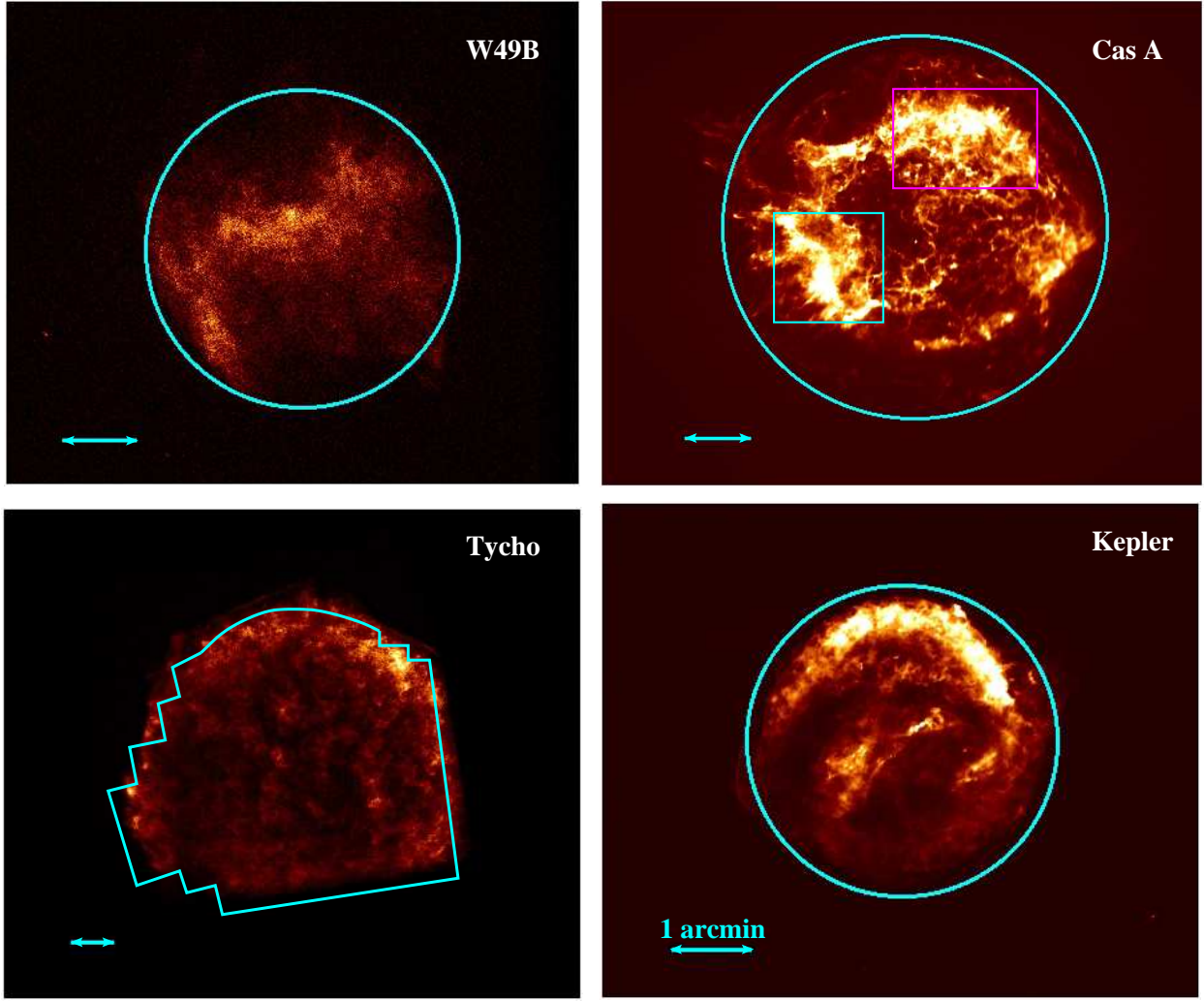}
\caption{{\sl Chandra} images of W49B, Cas A, Tycho and Kepler. The
source spectral regions are overplotted. In order to cover the whole
Tycho SNR, we use multi-regions as shown in the stretch. The
double-arrowed lines represent 1 arcmin for the corresponding image.
The two boxes in the Cas A image represent the blueshift (southeast,
blue one) and redshift (northwest, red one) dominated portions we
selected to create the spectra as described in $\S$ 3.2. }
\end{figure*}

\begin{figure*}
\includegraphics[width=\textwidth,clip]{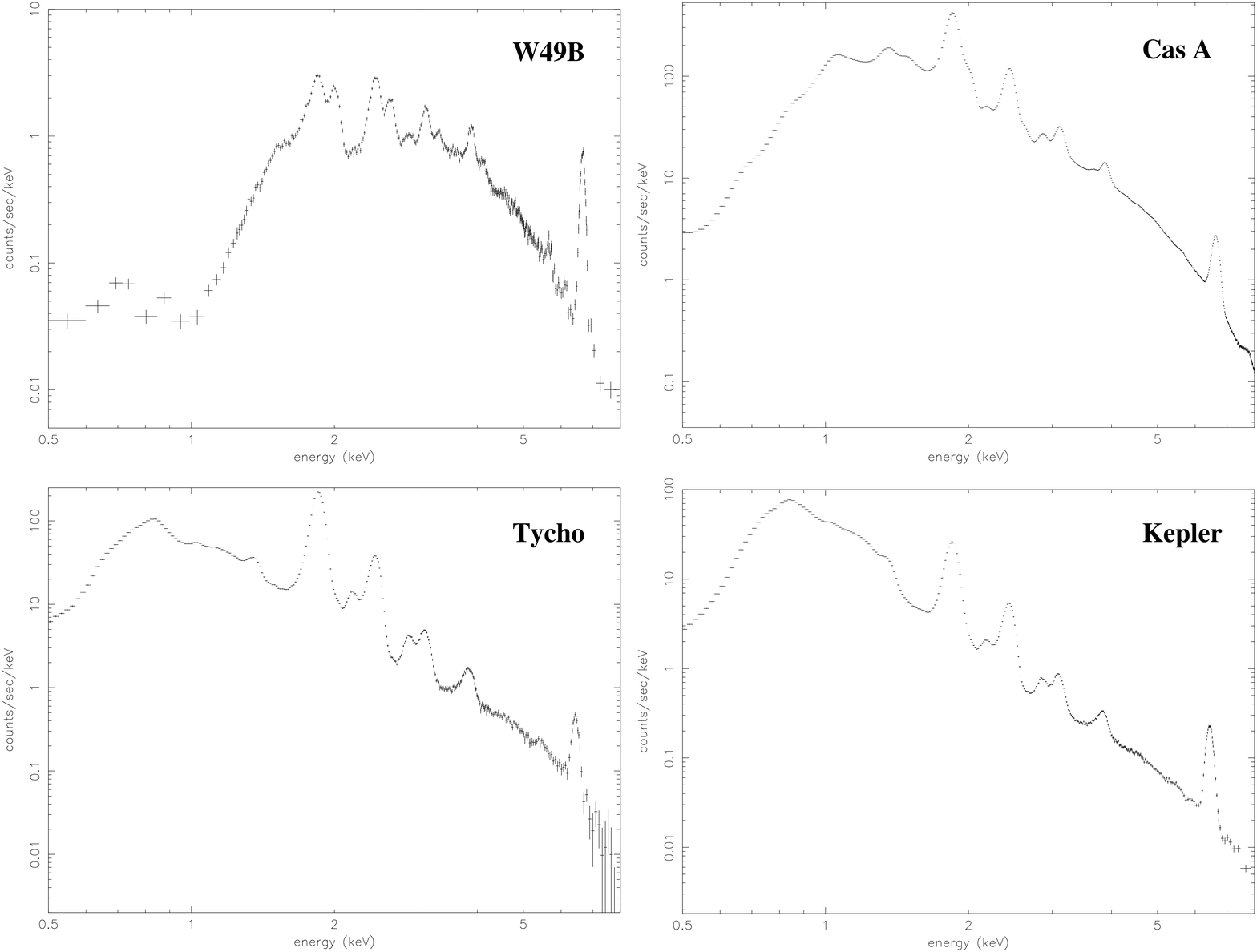}
\caption{The 0.5$-$8.0 keV spectra of W49B, Cas A, Tycho and Kepler.}
\end{figure*}

\begin{figure*}
\includegraphics[width=\textwidth,clip]{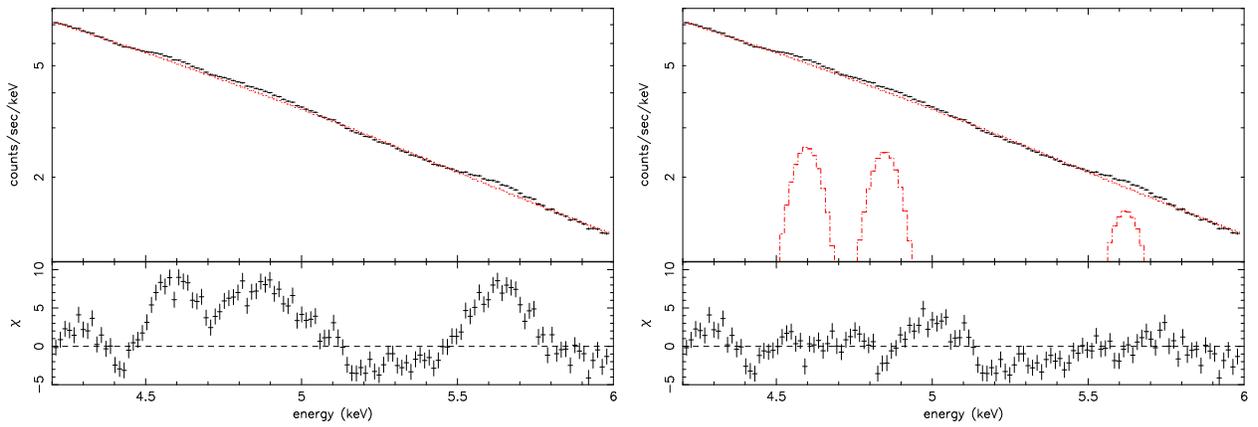}
\caption{The 4.2$-$6.0 keV spectrum of Cas A fitting with different
models. The left panel presents the fitting with one power law,
while the right panel one power law plus three Gaussian components.
The residual (normalized by sigma, $\chi$) distributions are also
plotted. The strength of all the Gaussian components are multiplied by a
factor of 15, so as to be shown clearly. }
\end{figure*}

\begin{figure*}
\includegraphics[width=\textwidth,clip]{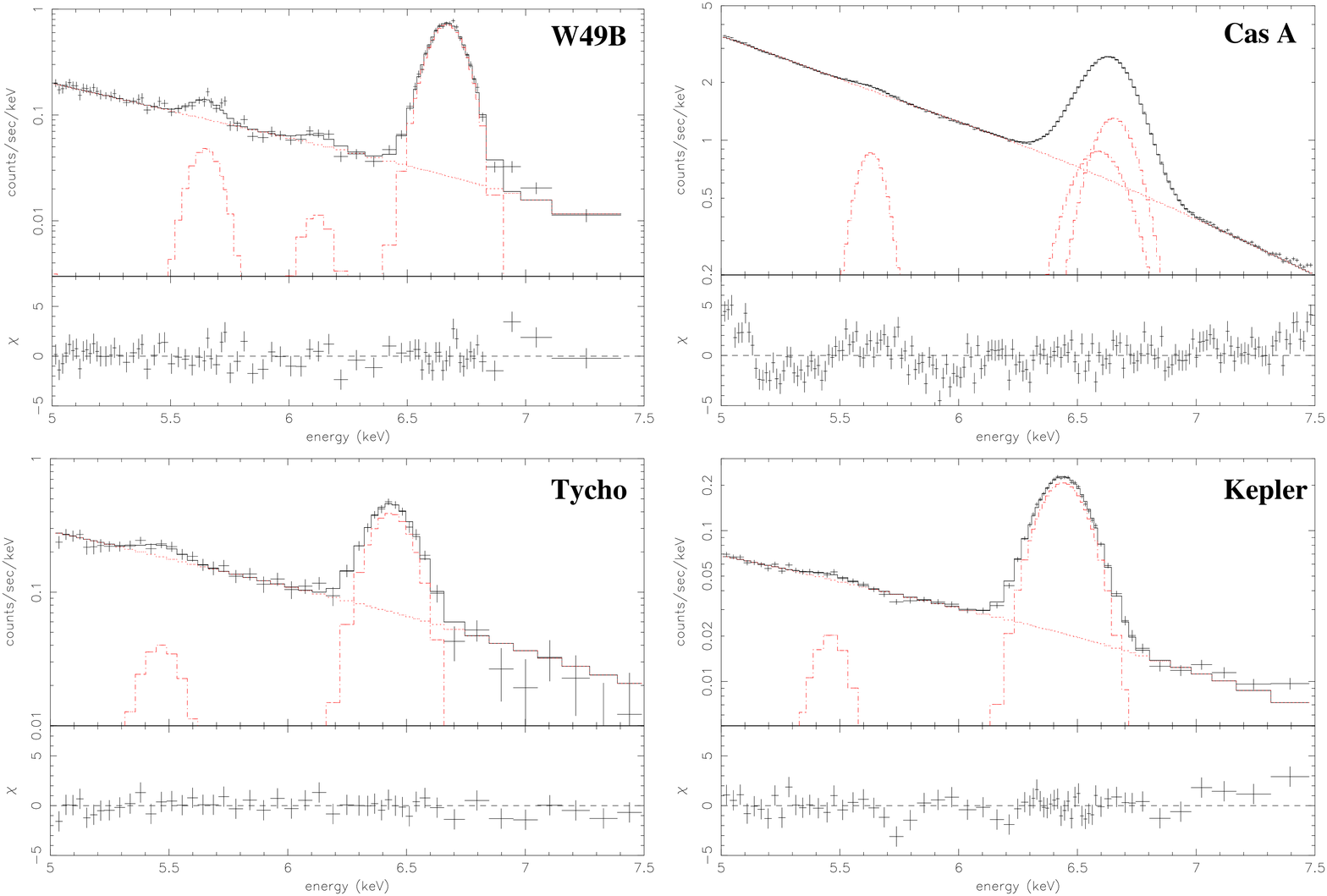}
\caption{The 5.0$-$7.5 keV spectra of W49B, Cas A, Tycho and Kepler.
The lines represent the fitting models as described in $\S$ 3. The
strength of the Cr line for Cas A is multiplied by a factor of 10,
while that for Kepler a factor of 5. }
\end{figure*}

\begin{figure*}
\includegraphics[width=8cm,angle=270,clip]{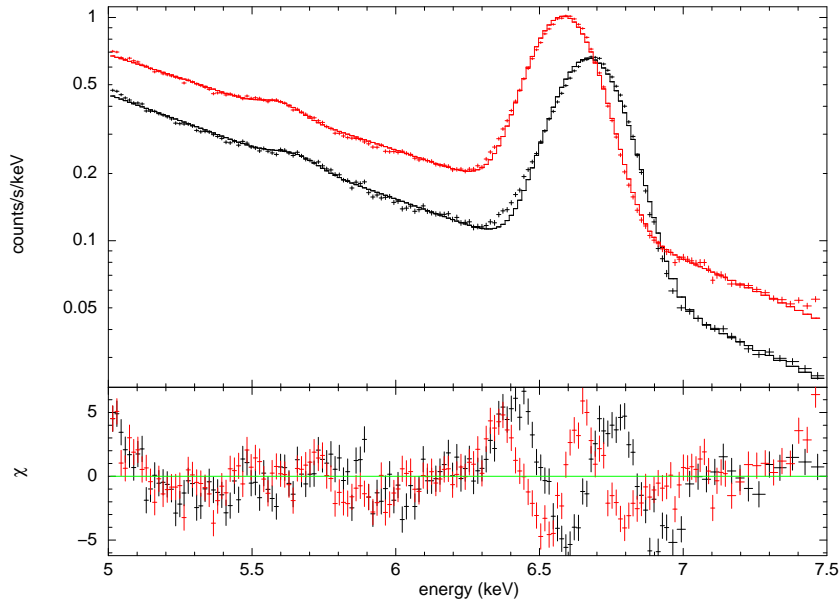}
\caption{The 5.0$-$7.5 keV spectra of Cas A from the redshift (upper
curve) and blueshift (lower curve) dominated portions. They are both
fitted with a model involving a power law plus two Gaussian components. }
\end{figure*}

\begin{figure*}
\includegraphics[width=12cm,clip]{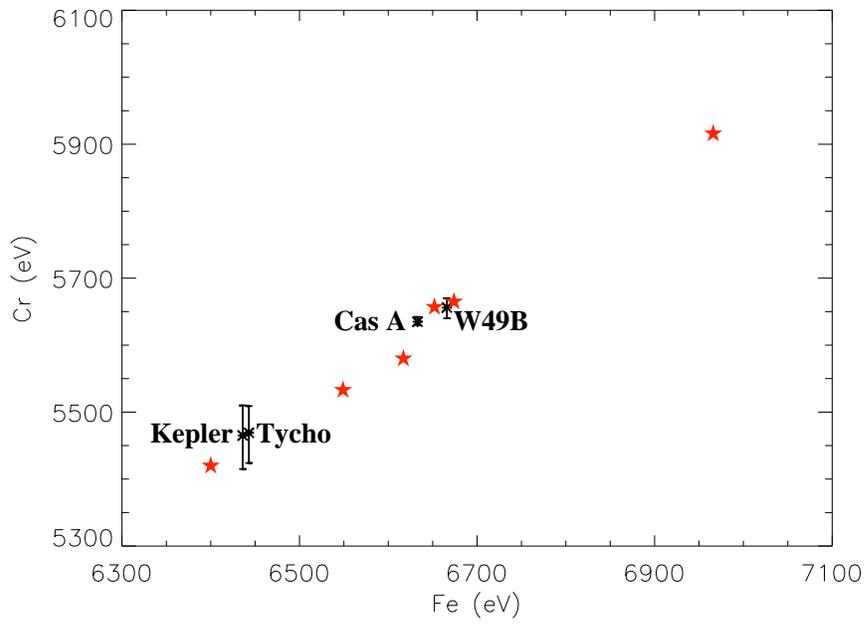}
\caption{The line center energy of Cr versus Fe in W49B, CasA, Tycho
and Kepler. The errors of the Fe line are not plotted, since they
are almost within the size of the symbols. The stars represent
the line center energy of Cr and Fe in various ionization states, as
given in Table 4. }
\end{figure*}

\end{document}